\documentclass[doublecol]{epl2}
\usepackage{graphicx,amsmath,amssymb,amsfonts,latexsym,color,dcolumn,bm,epsfig,subfigure}
\usepackage[latin1]{inputenc}

\def\bbm[#1]{\mbox{\boldmath $#1$}}
\newcommand{\ket}[1]{\displaystyle{|#1\rangle}}
\newcommand{\bra}[1]{\displaystyle{\langle #1|}}

\newcommand{\TE}{\text{TE}}
\newcommand{\TM}{\text{TM}}

\title{Steady entanglement out of thermal equilibrium}
\shorttitle{Steady entanglement out of thermal equilibrium} 

\author{Bruno Bellomo\inst{1,2} \thanks{E-mail: \email{bruno.bellomo@univ-montp2.fr}}  \and Mauro Antezza\inst{1,2,3} \thanks{E-mail: \email{mauro.antezza@univ-montp2.fr}}}
\shortauthor{B. Bellomo \etal}

\institute{
  \inst{1} Universit\'{e} Montpellier 2, Laboratoire Charles Coulomb UMR 5221 - F-34095, Montpellier, France, EU\\
  \inst{2} CNRS, Laboratoire Charles Coulomb UMR 5221 - F-34095, Montpellier, France, EU\\
   \inst{3} Institut Universitaire de France - 103, bd Saint-Michel
F-75005 Paris, France, EU
}


\abstract{We study two two-level atomic quantum systems (qubits) placed close to a body held at a temperature different from that of the surrounding walls. While at thermal equilibrium the two-qubit dynamics  is characterized by not entangled steady thermal states, we show that absence of thermal equilibrium may bring to the generation of entangled steady states. Remarkably, this entanglement emerges from the two-qubit dissipative dynamic itself, without any further external action on the two qubits, suggesting a new protocol to produce and protect entanglement which is intrinsically robust to environmental effects.}

\begin{document}

\maketitle

\section{Introduction}

Entanglement represents one of the key features in quantum mechanics \cite{Horodecki09}  due to its  connection to non locality \cite{Einstein35,Clauser69} and  its crucial role in quantum information \cite{BookNielsen}. Environmental noise \cite{BookBreuer} induces decoherence  \cite{Zurek03} and is typically responsible for the fragility of entanglement \cite{Yu04}. This represents one of the major obstacles to the concrete realization of quantum  technologies related to quantum information processing \cite{BookNielsen,Horodecki09}. A huge effort has been dedicated to the comprehension of the detrimental environmental effects \cite{Yu04, BellomoPRL07,BellomoPRA07, BellomoPRA12,Almeida07,Guo10} and in conceiving suitable approaches to contrast the natural decay of quantum correlations  \cite{Cirac09}.
 They include reservoir engineering \cite{Cirac09}, feedback methods \cite{Carvalho11}, distillation protocols \cite{Gisin01}, decoherence free-subspaces \cite{Lidar98}, non-Markovian effects \cite{BellomoPRL07}, weak measurements \cite{Kim12}, quantum Zeno effect \cite{ManiscalcoPRL08}, dynamical decoupling \cite{Viola99} and reservoir monitoring \cite{CarvalhoNJP11}.
Different protocols exploiting dissipative effects to realize steady  entanglement have been proposed  \cite{Plenio02,Hartmann06,Krauter11}.

Here, we introduce a  direct procedure to protect entanglement  realized by bringing the environment of a two-qubit system out of thermal equilibrium.
Physical systems consisting of two qubits in a common environment in absence \cite{ Braun02, Benatti03, Tanas08} or presence \cite{Tudela11, Xu11} of matter have been largely investigated at thermal equilibrium, pointing out the creation of entanglement due to the field mediated interaction, which however typically washes off asymptotically.
Efforts have been also done considering two or more qubits interacting with independent thermal reservoirs at different temperatures, pointing out  the possible creation of steady entangled states \cite{Quiroga07,Huang09,Camalet2011,Wu11,Znidaric12,Camalet2013}. Thermal reservoirs at distinct temperatures are also exploited in thermal machines involving few atoms \cite{Linden11}.  On the other hand,
new possibilities emerging in realistic systems out of thermal equilibrium, keeping into account the scattering matrices of the bodies present in the system, have been recently pointed out in different contexts ranging form heat transfer \cite{Joulain, Messina12}, to Casimir-Lifshits forces \cite{AntezzaPRL05,AntezzaJPA06, ObrechtPRL07, AntezzaPRA08,Bimonte09,KardarPRL2011,MesAntEPL11,MesAntPRA11} and atomic dynamics \cite{BellomoEPL2012,BellomoPRA2013}. In particular, near field effects in the case of atoms close to bodies are relevant.

In this Letter we investigate how entanglement between two qubits  can be manipulated by means of a complex electromagnetic field out of thermal equilibrium resulting from the presence of bodies at different temperature  whose geometrical and dielectric properties can be used as a resource.
We will show that this  environmental noise has two remarkable effects: it \emph{contrasts} the usual dechoerence between the qubits, and it \emph{generates} steady entangled states. This  is obtained without any further external actions on the two qubits, such as the use of lasers or complex procedures involving measurements on the qubits or on the environment. Differently from \cite{Camalet2011}, the main effects emerging out of equilibrium are here obtained by means of a  \emph{single} common field.

\begin{figure}[h!]
\hspace{0.4cm}\includegraphics[width=0.465\textwidth]{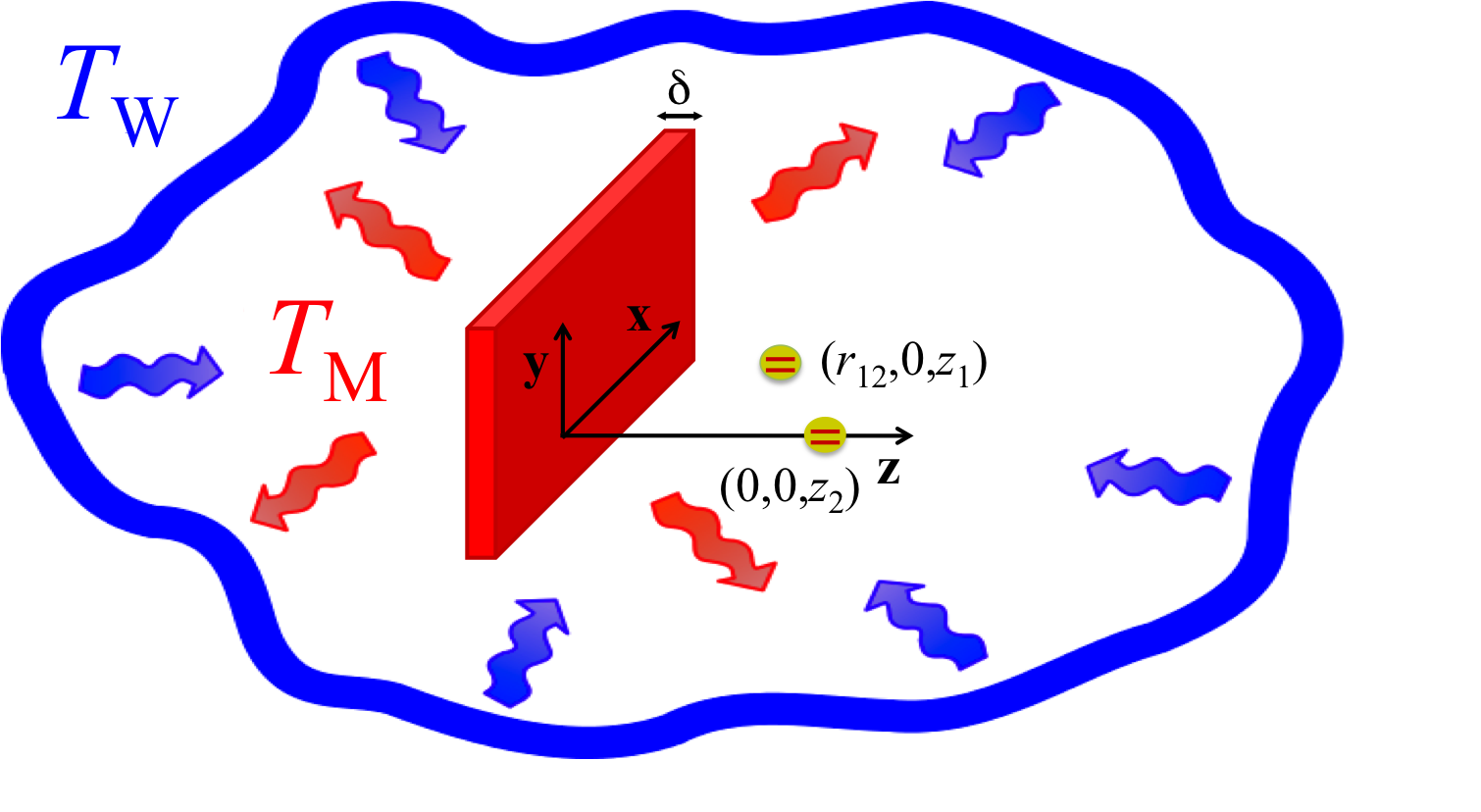}
\caption{\label{fig:1}\footnotesize  (color online). Two qubits close to a slab at temperature $T_\mathrm{M}$ different from the temperature  of the  surrounding walls, $T_\mathrm{W}$. The two qubits are placed in $(\mathbf{r}_1, z_1)$ and $(\mathbf{r}_2, z_2)$, where $\mathbf{r}_1$ and $\mathbf{r}_2$ are vectors in the $xy$ plane and $r_{12}=|\mathbf{r}_1-\mathbf{r}_2|$.}
\end{figure}
\textit{Physical system and model.\textemdash}We consider two qubits $q=1, 2$, whose ground $\ket{g}_q$ and excited $\ket{e}_q$ internal levels
are separated by the frequency $\omega=\omega_e^1-\omega_g^1=\omega_e^2-\omega_g^2$, interacting with a complex environment consisting in a stationary out of thermal equilibrium electromagnetic field. This is the result of the field emitted by a body of arbitrary geometry and dielectric permittivity, held at the temperature $T_\mathrm{M}$, and of the field emitted by far surrounding walls held at temperature $T_\mathrm{W}$, eventually reflected and transmitted by the body (see Fig. \ref{fig:1} when the body is a slab).
The walls have an irregular shape and are distant enough from the qubits such that the field they would produce at the qubits position (in the absence of the body) would be a blackbody radiation independent from the walls composition and geometry \cite{AntezzaPRL05,AntezzaJPA06, ObrechtPRL07, AntezzaPRA08}. The total Hamiltonian has the form $H=H_S+H_E+H_I$,
where $H_S= \sum_q\sum_{ n={g,e}} \hbar \omega_{n}^q \sigma_{nn}^q$, being  $\sigma_{mn}^q=\ket{m}_{qq}\bra{n}$, is the free two-qubit Hamiltonian and $H_E$ the free environmental Hamiltonian.
The interaction between the qubits and the environment in the multipolar coupling and in dipole approximation is described by
$H_I=-\sum_q \mathbf{D}_q\cdot\mathbf{E}(\mathbf{R}_q) $ \cite{CohenTannoudji97},
where $\mathbf{D}_q$  is the electric-dipole operator of qubit $q$ (being ${}_q\bra{g}\mathbf{D}_q\ket{e}_q=\mathbf{d}^q$), and $\mathbf{E}(\mathbf{R}_q) $ is the electric field at the position $\mathbf{R}_q$ of qubit $q$.

\section{Master equation}

The starting point to study the two-qubit dynamics is the von Neumann equation for the total density matrix,  which in the interaction picture is
$\dot{\rho}_{\text{tot}}(t)=-\frac{i}{\hbar}[H_I(t),\rho_{\text{tot}}(t)]$.
By tracing over the environmental degrees of freedom, after the Born, Markov and rotating wave approximations, the master equation for the reduced two-qubit density matrix becomes \cite{Agarwal1974,FicekBook2005}
\begin{equation}\begin{split}\label{master equation 6}&\frac{d}{d t}\rho =-\frac{i}{\hbar} [H_S+  \delta_S,\rho]
-i\sum_{q\neq q'}\Lambda^{qq'}(\omega)[\sigma_{ge}^{q\,\dag}\sigma_{ge}^{q'},\rho ] \\&
+\sum_{q, q'}\Gamma^{qq'}(\omega)\Big(\sigma^{q'}_{ge}\rho \sigma_{ge}^{q\,\dag}-\frac{1}{2}\{\sigma^{q\,\dag}_{ge}\sigma^{q'}_{ge},\rho \}\Big)\\&+\sum_{q, q'}\Gamma^{qq'}(-\omega)\Big(\sigma^{q'\dag}_{ge}\rho \sigma^{q}_{ge}-\frac{1}{2}\{\sigma^{q}_{ge} \sigma^{q'\,\dag}_{ge},\rho \}\Big),\end{split}\end{equation}
where $\delta_S$ is an operator related to the level frequency shifts, not playing any role in the following.
Function $\Lambda^{qq'}(\omega)$ represents temperature independent induced coherent (dipole-dipole) interaction between
the qubits, while $\Gamma^{qq'}(\pm\omega)$ are individual ($q=q'$) and common field mediated collective ($q\neq q'$) qubit transition rates, related to both quantum and thermal fluctuations of the electromagnetic field at the qubits positions.

In the following, we will use two  different basis:
the decoupled bases  $\{\ket{1}\equiv \ket{g g},\ket{2}\equiv \ket{e g},\ket{3}\equiv \ket{g e},\ket{4}\equiv \ket{e e}\}$, and the coupled bases $\{\ket{\mathrm{G}}\equiv \ket{1},
\ket{\mathrm{A}}\equiv (\ket{2}-\ket{3})/\sqrt{2},\ket{\mathrm{S}}\equiv(\ket{2}+\ket{3})/\sqrt{2},\ket{\mathrm{E}}\equiv \ket{4}\}$, where the  collective anti-symmetrical and symmetrical states $\ket{\mathrm{A}}$ and $\ket{\mathrm{S}}$ are combinations of the decoupled states $\ket{2}$ and $\ket{3}$.

\section{X states and concurrence}

In the decoupled basis, master equation \eqref{master equation 6} implies that the dynamics of the elements along the two main diagonals of the two-qubit density matrix (forming an  X-structure) is independent from that of the remaining ones.
Then, an initial state with an  X-structure maintains its form in time. Moreover, terms outside the two main diagonals, are washed off asymptotically.
 Bell, Werner and Bell diagonal states belong to the class of X states
\cite{BellomoASL}, which
arise in a wide variety of physical situations and are experimentally achievable \cite{Pratt2004PRL}.  In the following we will deal with X states.

We quantify the two-qubit entanglement  by means of the
concurrence $C(t)$
($C=0$ for separable states, $C=1$ for  maximally entangled states)~\cite{wootters1998PRL}.
For  X states, using $\rho_{ij}=\bra{i}\rho\ket{j}$, it takes the simple form~\cite{Yu2007}
\begin{equation}\begin{split}\label{concxstate}
C(t)&=2\;\mathrm{max}\{0,K_1(t),K_2(t)\}, \\
K_1(t)&=|\rho_{23}(t)|-\sqrt{\rho_{11}(t)\rho_{44}(t)},\\
K_2(t)&=|\rho_{14}(t)|-\sqrt{\rho_{22}(t)\rho_{33}(t)}.
\end{split}\end{equation}
Eq. \eqref{master equation 6} induces an exponential decay for $\rho_{14}(t)$, so that in the steady state only $K_1(t)$ could lead to $C(\infty)>0$.

\section{Thermal equilibrium}

For $T_\mathrm{W}=T_\mathrm{M}$ master equation \eqref{master equation 6} describes the qubits thermalization towards the diagonal thermal equilibrium state\footnote{The thermal state is not reached asymptotically if $\Gamma^{ii}(\pm\omega)=\Gamma^{ij}(\pm\omega)$. In this case, both at and out  equilibrium, the steady state depends on the initial state.}
\begin{equation}\label{thermal state}
  \left(
\begin{array}{c}
 \rho_{11}(\infty)\\
    \rho_{22}(\infty)\\
  \rho_{33}(\infty)\\
    \rho_{44}(\infty)\\
\end{array}
\right)_{\!\!\!\mathrm{eq}} =\frac{1}{Z_{\mathrm{eq}}}\left(
\begin{array}{c}
 [1+ n(\omega, T)]^2 \\
   n(\omega, T)[1+ n(\omega, T)] \\
 n(\omega, T)[1+ n(\omega, T)]\\
   n(\omega, T)^2\\
\end{array}
\right),
\end{equation}
where $Z_{\mathrm{eq}}=[1+2\;n(\omega, T)]^2$ and $n(\omega, T)=(\mathrm{e}^{\frac{\hbar\omega}{k_\mathrm{B}T}}-1)^{-1}$.
This state is universal, it depends only on the ratio $\hbar\omega/k_\mathrm{B}T$, remaining insensible to all system details.
Being $|\rho_{23}(\infty)|=0$, $K_1(\infty)$ is always negative, resulting in \emph{not entangled steady states}.  In terms of the density matrix in the coupled bases, using $\rho_{\mathrm{X}}\equiv\rho_{\mathrm{XX}}=\bra{\mathrm{X}}\rho\ket{\mathrm{X}}$,
$\rho_{23}$ is equal to zero in the steady state since $\rho_{\mathrm{AS}}(\infty)=0$ and $\rho_{\mathrm{S}}(\infty)=\rho_{\mathrm{A}}(\infty)$.
The latter identity has not to be valid out of thermal equilibrium, allowing  $K_1(\infty)>0$ hence producing steady entanglement.

\section{Out of thermal equilibrium}

For $T_\mathrm{W}\neq T_\mathrm{M}$, the analysis of Eqs. (\ref{master equation 6}-\ref{concxstate}) is much more rich and delicate. The $\Gamma$ and $\Lambda$ functions depend on the correlation functions of the electromagnetic field, which at thermal equilibrium can be directly evaluated exploiting the fluctuation-dissipation theorem (FDT). Out equilibrium, the FTD is not valid in general. Nevertheless, we assume that the radiation emission by the body and the walls has the same characteristics it would have at thermal equilibrium at the source temperature \cite{AntezzaPRL05, AntezzaJPA06, ObrechtPRL07, AntezzaPRA08, MesAntEPL11, MesAntPRA11}. This allows to compute the correlation functions by indirectly using the FDT, as recently used to study  the dynamics of a single atom \cite{BellomoEPL2012,BellomoPRA2013}. The transition rates in Eq. \eqref{master equation 6} can be set under the  form
\begin{equation}\begin{split}
\label{gamma functions finale}
\Gamma^{qq'}(\omega)=& \sqrt{\Gamma_0^q(\omega)\Gamma_0^{q'}(\omega) } \Big\{[1+n(\omega,T_\mathrm{W})]\alpha_\mathrm{W}^{q q'}(\omega)
\\& + [1+n(\omega,T_\mathrm{M})]\alpha_\mathrm{M}^{q q'}(\omega)\Big\}\\
\Gamma^{qq'}(-\omega)= & \sqrt{\Gamma_0^q(\omega)\Gamma_0^{q'}(\omega) } \Big\{n(\omega,T_\mathrm{W})  \alpha_\mathrm{W}^{q q'}(\omega)^*
\\&+n(\omega,T_\mathrm{M})]\alpha_\mathrm{M}^{q q'}(\omega)^*\Big\}
 ,\end{split}\end{equation}
where
$
\alpha_\mathrm{W}^{q q'}(\omega) =\sum_{i,i'}[\tilde{\textbf{d}}^q]^*_{i} [\tilde{\textbf{d}}^{q'}]_{i'} [\alpha_\mathrm{W}^{q q'}(\omega)]_{ii'} $, $
\alpha_\mathrm{M}^{q q'}(\omega) =\sum_{i,i'}[\tilde{\textbf{d}}^q]^*_{i} [\tilde{\textbf{d}}^{q'}]_{i'} [\alpha_\mathrm{M}^{q q'}(\omega)]_{ii'}$,
being $[\tilde{\textbf{d}}^q]_{i}=[\textbf{d}^q]_{i}/|\textbf{d}^q|$,  and $\Gamma_0^q(\omega)=|\textbf{d}^q|^2\omega^3/3  \hbar  \pi\epsilon_0 c^3 $ is the vacuum spontaneous-emission rate of qubit $q$.  $[\alpha_\mathrm{M}^{q q'}(\omega)]_{ii'} $ and $[\alpha_\mathrm{W}^{q q'}(\omega)]_{ii'} $ are temperature independent functions, which depend on all the other system parameters  and can be expressed as
\begin{equation}\label{alphaWM}\begin{split}
&[\alpha_\mathrm{W}^{q q'}(\omega)]_{ii'}=\frac{3\pi c}{2 \omega}\sum_{p,p'}\int\frac{d^2\mathbf{k}}{(2\pi)^2}\int\frac{d^2\mathbf{k}'}{(2\pi)^2}e^{i(\mathbf{k}\cdot\mathbf{r}_q-\mathbf{k}'\cdot\mathbf{r}_{q'})}\\ &
\, \times \bra{p,\mathbf{k}}\Bigl\{e^{i(k_z z_q-k_z^{'*}z_{q'})}[\hat{\bbm[\epsilon]}_p^+(\mathbf{k},\omega)]_i[\hat{\bbm[\epsilon]}_{p'}^{+}(\mathbf{k}',\omega)]_{i'}^*\\
&\,\times \Bigl(\mathcal{T}\mathcal{P}_{-1}^{\text{(pw)}}\mathcal{T}^{\dag}+\mathcal{R}\mathcal{P}_{-1}^{\text{(pw)}}\mathcal{R}^{\dag}\Bigr)+e^{i(k_zz_q+k_z^{'*}z_{q'})} \\&
\, \times [\hat{\bbm[\epsilon]}_p^+(\mathbf{k},\omega)]_i[\hat{\bbm[\epsilon]}_{p'}^{-}(\mathbf{k}',\omega)]_{i'}^*\mathcal{R}\mathcal{P}_{-1}^{\text{(pw)}}+e^{-i(k_zz_q+k_z^{'*}z_{q'})}\\
&  \, \times[\hat{\bbm[\epsilon]}_p^-(\mathbf{k},\omega)]_i[\hat{\bbm[\epsilon]}_{p'}^{+}(\mathbf{k}',\omega)]_{i'}^*\mathcal{P}_{-1}^{\text{(pw)}}\mathcal{R}^{\dag}+e^{-i(k_z z_{q}-k_z^{'*}z_{q'})} \\& \, \times [\hat{\bbm[\epsilon]}_p^-(\mathbf{k},\omega)]_i[\hat{\bbm[\epsilon]}_{p'}^{-}(\mathbf{k}',\omega)]_{i'}^*\mathcal{P}_{-1}^{\text{(pw)}}\Big\}\ket{p',\mathbf{k}'} ,
\\
&[\alpha_\mathrm{M}^{q q'}(\omega)]_{ii'}=\frac{3\pi c}{2 \omega}\sum_{p,p'}\int\frac{d^2\mathbf{k}}{(2\pi)^2}\int\frac{d^2\mathbf{k}'}{(2\pi)^2}e^{i(\mathbf{k}\cdot\mathbf{r}_q-\mathbf{k}'\cdot\mathbf{r}_{q'})}  \bra{p,\mathbf{k}}
\\ &  \, \Bigl\{e^{i(k_z z_q-k_z^{'*}z_{q'})}[\hat{\bbm[\epsilon]}_p^+(\mathbf{k},\omega)]_i[\hat{\bbm[\epsilon]}_{p'}^{+}(\mathbf{k}',\omega)]_{i'}^*\Bigl[\Bigl(\mathcal{P}_{-1}^\text{(pw)} +\mathcal{R}\mathcal{P}_{-1}^\text{(ew)}
\\
&\, -\mathcal{P}_{-1}^\text{(ew)}\mathcal{R}^{\dag} -\mathcal{R}\mathcal{P}_{-1}^\text{(pw)}\mathcal{R}^{\dag}  -\mathcal{T}\mathcal{P}_{-1}^\text{(pw)}\mathcal{T}^{\dag}\Bigr)\,\Big\}\ket{p',\mathbf{k}'},\end{split}\end{equation}
where the operators $\mathcal{R}$ and $\mathcal{T}$ are the standard reflection and transmission scattering operators, explicitly defined for example in \cite{MesAntPRA11}, associated in this case to the right side of the body. They connect any outgoing (reflected or transmitted) mode of the field to the entire set of incoming modes.
In the previous equations, each mode of the field is identified by the frequency $\omega$, the transverse wave vector $\mathbf{k}=(k_x,k_y)$, the polarization index $p$ (taking the values $p=1,2$ corresponding to TE and TM polarizations respectively), and the direction or propagation $\phi=\pm1$ (shorthand notation $\phi=\pm$) along the $z$ axis. In this approach, the total wavevector takes the form $\mathbf{K}^\phi=(\mathbf{k},\phi k_z)$, where the $z$ component of the wavevector $k_z$ is a dependent variable given by $k_z=\sqrt{\frac{\omega^2}{c^2}-k^2}$, where $k=|\mathbf{k}|$.
For the polarization vectors appearing in Eq. \eqref{alphaWM}  we adopt the following standard definitions
$
\hat{\bbm[\epsilon]}^\phi_\TE(\mathbf{k},\omega)=\hat{\mathbf{z}}\times\hat{\mathbf{k}}=(-k_y\hat{\mathbf{x}}+k_x\hat{\mathbf{y}})/k,
\hat{\bbm[\epsilon]}^\phi_\TM(\mathbf{k},\omega)=c\, \hat{\bbm[\epsilon]}^\phi_\TE(\mathbf{k},\omega)\times\mathbf{K}^{\phi}/\omega=c\, (-k\hat{\mathbf{z}}+\phi k_z\hat{\mathbf{k}})/\omega,
$
where $\hat{\mathbf{x}}$, $\hat{\mathbf{y}}$ and $\hat{\mathbf{z}}$ are the unit vectors along the three axes and $\hat{\mathbf{k}}=\mathbf{k}/k$. In Eq. \eqref{alphaWM} we have also used
$ \bra{p,\mathbf{k}}\mathcal{P}_n^\text{(pw/ew)}\ket{p',\mathbf{k}'}=k_z^n\bra{p,\mathbf{k}}\Pi^\text{(pw/ew)}\ket{p',\mathbf{k}'}$,
$\Pi^\text{(pw)}$ and $\Pi^\text{(ew)}$ being the projectors on the propagative ($ck<\omega$, corresponding to a real $k_z$) and evanescent ($ck>\omega$, corresponding to a purely imaginary $k_z$) sectors respectively.

With regards to the $\Lambda^{qq'}(\omega)$ function of Eq. \eqref{master equation 6}, it can be also expressed in terms of $\alpha$ functions as
\begin{equation}\label{lambda2}\begin{split}
&\Lambda^{qq'}(\omega)= \frac{\sqrt{\Gamma_0^q(\omega)\Gamma_0^{q'}(\omega) }}{\omega^3}
\\ & \times  \mathcal{P}\int_{-\infty}^{+\infty}\frac{\omega'^3 d\omega'}{2\pi}\frac{\alpha_\mathrm{W}^{q q'}(\omega') +\alpha_\mathrm{M}^{q q'}(\omega') }{\omega-\omega'}.
 \end{split}
 \end{equation}
 Previous expression can be calculated
 by  exploiting the connection between $\alpha$ functions and the imaginary part of the Green function of the system, $
 \mathrm{Im}\, G_{ii'} (\mathbf{R}_q,\mathbf{R}_{q'},\omega)=\frac{\omega^3}{3\pi \epsilon_0c^3}\frac{[\alpha_\mathrm{W}^{q q'}(\omega)]_{ii'} +[\alpha_\mathrm{M}^{q q'}(\omega)]_{ii'}}{2}$,
where $i$ and $i'$ refer to the cartesian components of the
field and $G_{ii'}(\mathbf{R}_q,\mathbf{R}_{q'},\omega)$ is the $ii'$
component of the Green function of the system.
Eq. \eqref{lambda2} thus becomes
\begin{equation}\label{lambda3}\begin{split}
\Lambda^{qq'}(\omega)=-\frac{1}{\hbar} \sum_{i,i'}  [\mathbf{d}^q]_i^* [\mathbf{d}^{q'}]_{i'}  \mathrm{Re} \, G_{ii'} (\mathbf{R}_q,\mathbf{R}_{q'},\omega),
 \end{split}
 \end{equation}
where Kramers-Kronig relations connecting real and imaginary parts of the green function have been used to compute the principal value of the integral. $\mathrm{Re} \, G_{ii'} (\mathbf{R}_q,\mathbf{R}_{q'},\omega)$ can then obtained after having derived the Green function of the system $G_{ii'} (\mathbf{R}_q,\mathbf{R}_{q'},\omega)$ \cite{Bellomopreparation}.

\begin{figure}[t!]
\includegraphics[width=0.47\textwidth]{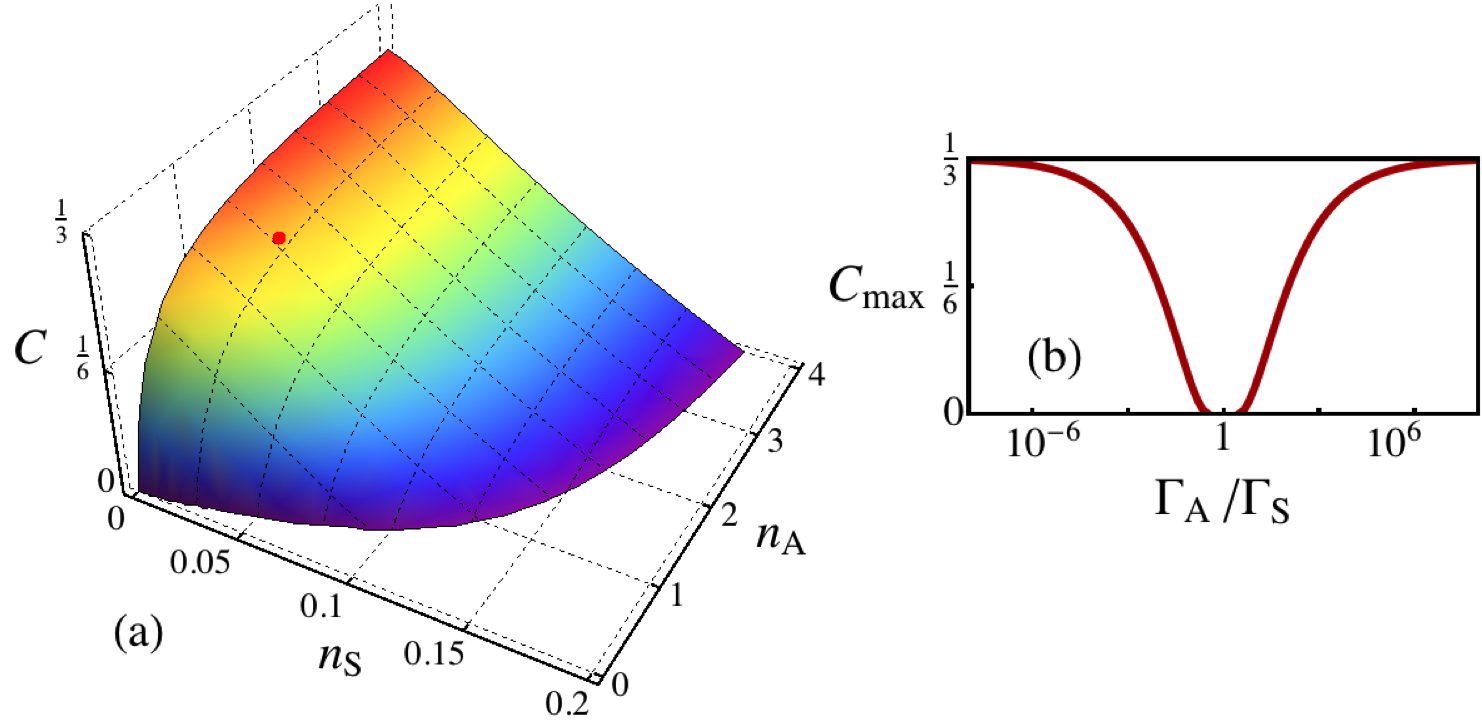}
\caption{\label{fig:figuraconcorrenza}\footnotesize (color online). Part (a): steady concurrence [$C=C(\infty)$] vs $n_\mathrm{S}$ and $n_\mathrm{A}$ for a fixed value $\Gamma_\mathrm{A}/\Gamma_\mathrm{S}\approx 2.8 \times 10^{-4} $. The red point corresponds to the maximum of concurrence along the white line in part (a) of Fig. \ref{fig:concorrenzacupola}. Part (b): maximum of concurrence, $C_{\mathrm{max}}$ as function of $\Gamma_\mathrm{A}/\Gamma_\mathrm{S}$.}
\end{figure}

\textit{Analytical investigation.\textemdash}To illustrate the new qualitative and quantitative behaviour of the entanglement out of thermal equilibrium, we first consider an instructive case allowing a direct interpretation. Let us consider $\Gamma^{11}(\pm\omega)=\Gamma^{22}(\pm\omega)\equiv \Gamma(\pm\omega)$ and $\Gamma^{12(21)}(\pm\omega) \in  \mathbb{R}$. These conditions are verified for identical qubits ($\mathbf{d}^1=\mathbf{d}^2\equiv\mathbf{d} $) in equivalent positions with respect to the body (in the case the body is a slab, $z_1=z_2$) and with $\mathbf{d}$ real and directed or along the $z$ axis or  along the $x-y$ plane. In this case,  master equation \eqref{master equation 6} implies in the coupled basis  a set of \emph{rate equations} for the populations, decoupled from the other density matrix elements:
\begin{figure}[t!]
\hspace{0.2cm}\includegraphics[width=0.468\textwidth]{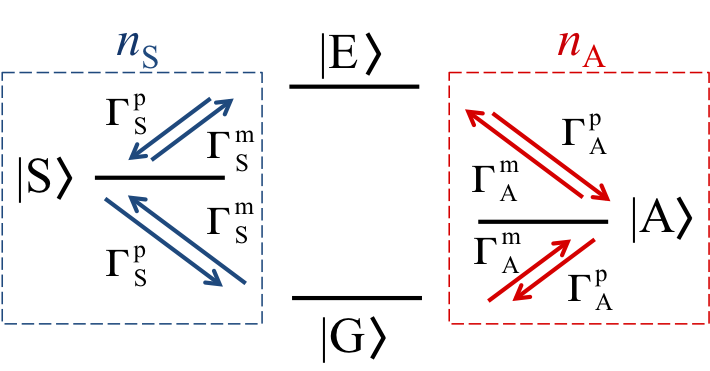}
\caption{\label{fig:figuraschema}\footnotesize (color online). Scheme of the rate equations of Eq. \eqref{MEsymm}: $\Gamma^p_{\mathrm{S}(\mathrm{A})}=\Gamma_{\mathrm{S}(\mathrm{A})}(1+ n_{\mathrm{S}(\mathrm{A})})$, $\Gamma^m_{\mathrm{S}(\mathrm{A})}=\Gamma_{\mathrm{S}(\mathrm{A})}\, n_{\mathrm{S}(\mathrm{A})}.$}
\end{figure}
\begin{equation}\begin{split}\label{MEsymm}
\dot{\rho}_\mathrm{G}=&\Gamma_\mathrm{A}(1+n_\mathrm{A}) \rho_\mathrm{A}+\Gamma_\mathrm{S}(1+n_\mathrm{S}) \rho_\mathrm{S}\\&-(\Gamma_\mathrm{A} \,n_\mathrm{A}+\Gamma_\mathrm{S}\, n_\mathrm{S})\rho_\mathrm{G}+,\\
\dot{\rho}_\mathrm{A}=&\Gamma_\mathrm{A}\, n_\mathrm{A} \rho_\mathrm{G}+\Gamma_\mathrm{A}(1+n_\mathrm{A}) \rho_\mathrm{E}-\Gamma_\mathrm{A} (1+2 n_\mathrm{A})  \rho_\mathrm{A},\\
\dot{\rho}_\mathrm{S}=&\Gamma_\mathrm{S}\, n_\mathrm{S} \rho_\mathrm{G}+\Gamma_\mathrm{S}(1+n_\mathrm{S}) \rho_\mathrm{E}-\Gamma_\mathrm{S} (1+2 n_\mathrm{S})  \rho_\mathrm{S},\\
\dot{\rho}_\mathrm{E}=&\Gamma_\mathrm{A}\, n_\mathrm{A} \rho_\mathrm{A} +\Gamma_\mathrm{S}\, n_\mathrm{S} \rho_\mathrm{S} \\& -[\Gamma_\mathrm{A} (1+ n_\mathrm{A}) +\Gamma_\mathrm{S} (1+ n_\mathrm{S}) ] \rho_\mathrm{E}.
\end{split}\end{equation}
Here the derivates are with respect to $\Gamma_0(\omega)t$ $[\Gamma_0(\omega)\equiv\Gamma_0^{(1)}(\omega)=\Gamma_0^{(2)}(\omega)]$,
and  we introduced the symmetric and anti-symmetric rates and the effective number of photons
\begin{equation}\begin{split}\label{nA e nS}
\Gamma_\mathrm{A}=&\alpha_\mathrm{W}(\omega) -\alpha^{12}_\mathrm{W}(\omega)+\alpha_\mathrm{M}(\omega) -\alpha^{12}_\mathrm{M}(\omega)\\
\Gamma_\mathrm{S}=&\alpha_\mathrm{W}(\omega) +\alpha^{12}_\mathrm{W}(\omega)+\alpha_\mathrm{M}(\omega) +\alpha^{12}_\mathrm{M}(\omega)\\
n_\mathrm{A}=&\frac{1}{\Gamma_\mathrm{A}}\Bigl\{ \left[\alpha_\mathrm{W}(\omega) -\alpha^{12}_\mathrm{W}(\omega)\right] n(\omega,T_\mathrm{W})
\\& +\left[\alpha_\mathrm{M}(\omega) -\alpha^{12}_\mathrm{M}(\omega)\right]n(\omega,T_\mathrm{M}) \Bigr\}\\
n_\mathrm{S}=& \frac{1}{\Gamma_\mathrm{S}}\Bigl\{ \left[\alpha_\mathrm{W}(\omega) +\alpha^{12}_\mathrm{W}(\omega)\right] n(\omega,T_\mathrm{W})
\\& +\left[\alpha_\mathrm{M}(\omega) +\alpha^{12}_\mathrm{M}(\omega)\right]n(\omega,T_\mathrm{M}) \Bigr\},
\end{split}\end{equation}
being $\alpha_\mathrm{W(M)}(\omega)\equiv\alpha^{11}_\mathrm{W(M)}(\omega)=\alpha^{22}_\mathrm{W(M)}(\omega)$.
Function $\Lambda$ does not enter in
the rate equations \eqref{MEsymm}, which are depicted in Fig. \ref{fig:figuraschema}. To each decay channel from $\ket{\mathrm{E}}$ to $\ket{\mathrm{G}}$, passing respectively trough $\ket{\mathrm{S}}$ and $\ket{\mathrm{A}}$, one can associate an effective number of photons $n_{\mathrm{S}(\mathrm{A})}$ confined between $n(\omega,T_\mathrm{W}) $ and $n(\omega,T_\mathrm{M})$, which is equivalent to associate an effective temperature $T_{\mathrm{S}(\mathrm{A})}$ confined between $T_\mathrm{W} $ and $T_\mathrm{M}$ \cite{BellomoEPL2012,BellomoPRA2013}.
While the coherences along the second diagonal decay exponentially to zero, the stationary solution of Eq. \eqref{MEsymm} is
\begin{widetext}
\begin{equation}\begin{split}\label{ote case}
\left(
\begin{array}{c}
 \rho_{\mathrm{G}}(\infty)\vspace{0.1 cm} \\
    \rho_{\mathrm{A}}(\infty)\vspace{0.1 cm} \\
   \rho_{\mathrm{S}}(\infty)\vspace{0.1 cm} \\
    \rho_{\mathrm{E}}(\infty)\\
\end{array}
\right)_{\!\!\!\mathrm{neq}} =
\frac{1}{Z_{\mathrm{neq}}}
\! \left(
\begin{array}{c}
\!\!(1+n_\mathrm{A})^2 (1+2n_\mathrm{S})\Gamma_\mathrm{A}+(1+2 n_\mathrm{A})(1+n_\mathrm{S})^2\Gamma_\mathrm{S} \!\! \vspace{0.1 cm} \\
n_\mathrm{A}(1+n_\mathrm{A})(1+2n_\mathrm{S}) \Gamma_\mathrm{A}+[ n_\mathrm{A}(1+2n_\mathrm{S}) +n_\mathrm{S}^2(1+2n_\mathrm{A}) ] \Gamma_\mathrm{S} \vspace{0.1 cm} \\
n_\mathrm{S}(1+n_\mathrm{S})(1+2n_\mathrm{A}) \Gamma_\mathrm{S}+[ n_\mathrm{S}(1+2n_\mathrm{A}) +n_\mathrm{A}^2(1+2n_\mathrm{S}) ] \Gamma_\mathrm{A} \vspace{0.1 cm} \\
n_\mathrm{A}^2 (1+2n_\mathrm{S})\Gamma_\mathrm{A}+(1+2 n_\mathrm{A})n_\mathrm{S}^2\Gamma_\mathrm{S}
\end{array}
\right)\!,
\end{split}\end{equation}
\end{widetext}
\begin{floatequation}
\mbox{\textit{see eq.~\eqref{ote case}}}
\end{floatequation}
where $Z_{\mathrm{neq}}$ is the sum of the elements of the vector on the right side of the above equation. Equation (\ref{ote case}), which reduces to the thermal state (\ref{thermal state}) for $T_{\mathrm{W}}=T_{\mathrm{M}}$, shows that out of equilibrium it is possible that $\rho_{\mathrm{S}}(\infty)\neq\rho_{\mathrm{A}}(\infty)$, and implies $  |\rho_{23}(\infty)|=|n_\mathrm{S}-n_\mathrm{A}|(\Gamma_\mathrm{S}+\Gamma_\mathrm{A})/2 Z_{\mathrm{neq}}$.  This leads to the possibility to have $K_1(\infty)>0$  in Eq. \eqref{concxstate}, corresponding to stationary entanglement.
Using Eq. \eqref{ote case} in Eq. \eqref{concxstate}, we obtain for the steady concurrence $C(\infty)=2\;\max\{0,K_1(\infty)\}$, with

\begin{equation}\begin{split}\label{concana}
& K_1(\infty)=\frac{1}{Z_{\mathrm{neq}}} \Bigl[ |n_\mathrm{S}-n_\mathrm{A}|(\Gamma_\mathrm{S}+\Gamma_\mathrm{A}) /2   \\
 &\,-\sqrt{(1+n_\mathrm{A})^2 (1+2n_\mathrm{S})\Gamma_\mathrm{A}+(1+2 n_\mathrm{A})(1+n_\mathrm{S})^2\Gamma_\mathrm{S}}\\
 &\, \times\sqrt{n_\mathrm{A}^2 (1+2n_\mathrm{S})\Gamma_\mathrm{A}+(1+2 n_\mathrm{A})n_\mathrm{S}^2\Gamma_\mathrm{S}}\Bigr],
\end{split}\end{equation}
which tends to zero at thermal equilibrium when $n_\mathrm{S}=n_\mathrm{A}$.
Simplifying $\Gamma_\mathrm{S}$, $C(\infty)$ becomes function of only $\Gamma_\mathrm{A}/\Gamma_\mathrm{S}$, $n_\mathrm{S}$ and $n_\mathrm{A}$. We discuss this dependence in Fig. \ref{fig:figuraconcorrenza} (a), where $C(\infty)$ is depicted as a function of $n_\mathrm{S}$ and $n_\mathrm{A}$ for $\Gamma_\mathrm{A}/\Gamma_\mathrm{S}\approx  2.8 \times 10^{-4} $. Large values of steady concurrence are obtained when the number of photons  associated to the two decay channels (see Fig.\ref{fig:figuraschema}) are enough distant between them. This physically corresponds to largely populate the antisymmetric state with respect to the symmetric one [see Eq. (\ref{ote case})]. By increasing too much  $n_\mathrm{A}$ at fixed  $n_\mathrm{S}$, the steady entanglement starts to decrease (not shown in the figure). Part (b) shows that the maximum value of $C(\infty)$ reachable by varying $n_\mathrm{S}$ and $n_\mathrm{A}$ at a fixed value of $\Gamma_\mathrm{A}/\Gamma_\mathrm{S}$ is $1/3$, obtainable in the two cases $\Gamma_\mathrm{A}/\Gamma_\mathrm{S}\to 0$ or $\Gamma_\mathrm{S}/\Gamma_\mathrm{A}\to 0$. The maximally entangled steady states are obtained in the first case in the limit of $n_\mathrm{S}\to 0$ and of $n_\mathrm{A}\to \infty$ while in the second case in the limit of  $n_\mathrm{A}\to 0$ and $n_\mathrm{S}\to \infty$. These states are a statistical mixture of the ground and of the antisymmetric (symmetric) state with weights respectively equal to 2/3 and 1/3 and are also found in \cite{Camalet2011}. We remark that up to now our findings do not rely on the specific choice of body's geometry or dielectric properties.

\section{Numerical investigation}

\begin{figure}[h!]
\includegraphics[width=0.47 \textwidth]{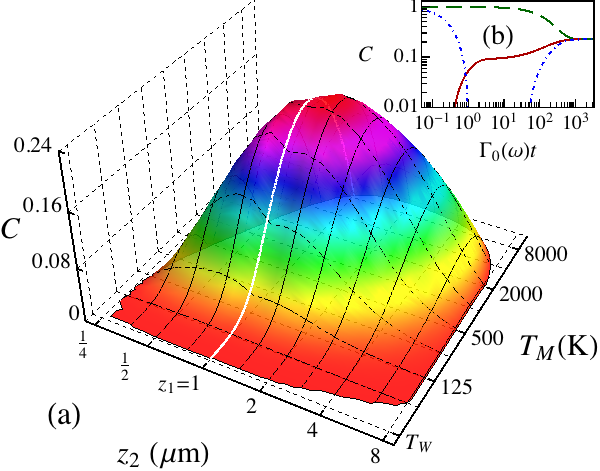}
\caption{\label{fig:concorrenzacupola}\footnotesize  (color online). Part (a): steady concurrence [$C=C(\infty)$] vs $z_2$ and $T_\mathrm{M}$. Here $z_1=1 \mu$m, $r_{12}=0.25 \mu$m (the qubits are distant $[r_{12}^2+(z_1-z_2)^2]^{1/2}$), $T_\mathrm{W}=30$ K, $\delta=0.01 \mu$m,   and $\omega=0.3  \omega_r$. Part (b):   concurrence as a function of $\Gamma_0(\omega)t$ for three different initial conditions, the antisymmetric (green dashed line), the symmetric (blue dotdashed line)  and the thermal state at $30$ K (red solid line).}
\end{figure}

In order to discuss the properties of $C(\infty)$ besides the case studied above,  we solve Eq. (\ref{master equation 6}) for the case where the body close to the two qubits is a slab of thickness $\delta$, as depicted in Fig. \ref{fig:1}.
In this case, we have at disposition simple expressions for $\mathcal{R}$ and $\mathcal{T}$. As a result of the translational invariance of a planar slab with respect to the $xy$ plane, its reflection and transmission operators, $\mathcal{R}$ and $\mathcal{T}$, are diagonal and given by
$\bra{p,\mathbf{k}}\mathcal{R}\ket{p',\mathbf{k}'}=(2\pi)^2\delta(\mathbf{k}-\mathbf{k}')\delta_{pp'}\rho_{p}(\mathbf{k},\omega)$ and $
\bra{p,\mathbf{k}}\mathcal{T}\ket{p',\mathbf{k}'}=(2\pi)^2\delta(\mathbf{k}-\mathbf{k}')\delta_{pp'}\tau_{p}(\mathbf{k},\omega)$,
where $\rho_{p}(\mathbf{k},\omega)$ and $\tau_{p}(\mathbf{k},\omega)$ are the Fresnel reflection and transmission coefficients of a slab of finite thickness $\delta$ \cite{BellomoPRA2013}.
It follows that  $\alpha_\mathrm{M}^{qq'}$ and $\alpha_\mathrm{W}^{qq'}$ of Eq. \eqref{alphaWM} reduce to simple integrals over propagative and evanescent sectors \cite{Bellomopreparation}. We also choose a SiC slab, describing its dielectric permittivity $\varepsilon(\omega)$ with a Drude-Lorentz model, with a resonance at $\omega_r=1.495\times10^{14}\,\mathrm{rad}\,\text{s}^{-1}$ and a surface {phonon-polariton} resonance at $\omega_p=1.787\times10^{14}\,\mathrm{rad}\,\text{s}^{-1}$. Hence, relevant length and temperature scales are $ c /\omega_r \simeq 2 \mu$m and $ \hbar \omega_r/k_B\simeq 1140$ K.
In Fig. \ref{fig:concorrenzacupola} (a) we plot concurrence of Eq. (\ref{concxstate}) as a function of $z_2$ and $T_\mathrm{M}$ in the case of two identical qubits having electric dipole perpendicular to the slab, for fixed values of $z_1=1\mu$m and $T_{\mathrm{W}}=30$ K.
The plot evidences a large zone in the space of the parameters corresponding to the generation of steady entangled states. The maximum value of $C(\infty)$, obtained for $z_1\neq z_2=1.28$ and $T_\mathrm{M}\approx 1300$ K, is $\approx 0.224$. The characteristic time to reach this entangled steady state is $\simeq  10^3 [\Gamma_0(\omega)]^{-1}$ [see part (b)]. The white line corresponds to the case $z_2=z_1$ and hence can be described by Eq. \eqref{concana}. The maximum along this curve, obtained for $T_\mathrm{M}\approx 1200$ K, corresponds to the red point in Fig. \ref{fig:figuraconcorrenza} (a), being  $n_\mathrm{A}\approx1.53$ $(T_\mathrm{A}\approx 680$ K) and $n_\mathrm{S}\approx 0.02$ $(T_\mathrm{S}\approx 90$ K). The relevant  difference between $T_\mathrm{S}$ and $T_\mathrm{A}$ is responsible of the high value of concurrence, $\approx 0.217$ (see also Fig. \ref{fig:figuraconcorrenza}). However,  by further increasing  $T_\mathrm{M}$  the concurrence decreases.  Values of steady concurrence higher than 0.14 are already obtained  at $T_\mathrm{M} \approx $ 500 K. High unphysical temperatures are here considered as an indication of what would occur at lower temperatures in the case of a slab made by a different material characterized by  similar values of  $\varepsilon(\omega)$ at lower frequencies.

In Fig. \ref{fig:concorrenzacupola} (b), by using the parameters corresponding to the maximum of Fig. \ref{fig:concorrenzacupola}(a), we show the time evolution of concurrence for three different initial states: the maximally entangled antisymmetric and symmetric states, and the not entangled thermal state at $T=30$ K. We note how the protection of entanglement and its steady production, respectively, are independent on the initial two-qubit state.
A systematic study shows also that by increasing the slab thickness $\delta$, or the value of $r_{12}$, or moving the atomic frequency $\omega$ towards the slab resonances, or changing the two qubits electric dipole orientations steady entanglement typically reduces. We observe that even a small amount of mixed state entanglement, here produced, could be then distilled into a pure entangled state \cite{horodecki}.
We remark that results similar to the ones discussed above could be found  in a different range of frequencies by considering a different material for the slab such that close values for the dielectric permittivity are found at different values of $\omega$.

A first direct realization of our two-qubit system is made by two-level atoms. Our study also applies to other kind of physical systems like quantum dots or superconducting qubits. In this case other possible additional sources of environmental noise should be taken into account, if they are strong enough to overcome the effects driven by the electromagnetic field.

\section{Conclusions}

We investigated the dynamics of two qubits interacting with a common stationary field out of thermal equilibrium.  We predicted the occurrence of steady entangled states not depending on the initial two-qubit state, consisting then in a creation and/or protection of entanglement according to the nature of the initial configuration. For a relevant class of parameters we derived an analytical expression for concurrence, and explained the entanglement production in terms of rate equations driven by two different effective  temperatures associated to the two decay channels governing the passage from the two-qubit excited state to the ground state. We numerically studied the case where the body close to the qubits is a slab, finding concurrence up to $\approx 1/4$.
While at thermal equilibrium the entanglement decays to zero faster if the temperature is increased,
the present strategy to create and/or protect entanglement can be realized, quite counterintuitively,  starting from a thermal equilibrium configuration and \emph{increasing} only one of the two temperatures of the system. To further increase the amount of steady entanglement, systematic studies exploiting different body's geometries are envisaged.

\acknowledgements {Authors  thank R. Messina for useful discussions and acknowledge financial support from the Julian Schwinger Foundation. M.A. is member of the LabEx NUMEV.}


\begin{thebibliography}{0}
\bibitem{Horodecki09}
 \Name{Horodecki R. \etal  }
    \REVIEW{Rev. Mod. Phys.}{81}{2009}{865}.

\bibitem{Einstein35}
    \Name{Einstein  A., Podolsky  B. \and N. Rosen.}
    \REVIEW{Phys. Rev.}{47}{1935}{777}.

\bibitem{Clauser69}
 \Name{Clauser J. F. ,  Horne M. A., Shimony A. \and Holt  R. A. .}
    \REVIEW{Phys. Rev. Lett.}{23}{880}{1969}.

\bibitem{BookNielsen}
    \Name{Nielsen M. A.\and Chuang I. L.}
    \Book{Quantum Computation and Quantum Information}
    \Publ{Cambridge Univ. Press}
    \Year{2000}

\bibitem{BookBreuer}
    \Name{Breuer H.-P. \and Petruccione F.}
    \Book{The Theory of Open Quantum Systems}
    \Publ{Oxford Univ. Press, New York}
    \Year{2002}


\bibitem{Zurek03}
    \Name{Zurek W. H.}
    \REVIEW{Rev. Mod. Phys.}{75}{2003}{715}.


\bibitem{Yu04}
    \Name{Yu T. \and Eberly J. H..}
    \REVIEW{Phys. Rev. Lett.}{93}{2004}{140404}.

\bibitem{BellomoPRL07}
    \Name{Bellomo B. , Lo Franco R.  \and Compagno G.}
    \REVIEW{Phys. Rev. Lett.}{99}{2007}{160502}.

\bibitem{BellomoPRA07}
    \Name{Bellomo B. , Lo Franco R.  \and Compagno G.}
     \REVIEW{Phys. Rev. A}{77}{2008}{032342}.

\bibitem{BellomoPRA12}
    \Name{Lo Franco R. , Bellomo B., Andersson E.  \and  Compagno G.}
    \REVIEW{Phys. Rev. A}{85}{2012}{032318}.

\bibitem{Almeida07}
    \Name{Almeida M. P.  \etal}
    \REVIEW{Science}{316}{2007}{579582}.

\bibitem{Guo10}
    \Name{Xu J. S. \etal}
    \REVIEW{Phys. Rev. Lett.}{104}{2010}{100502}.

\bibitem{Cirac09}
    \Name{Verstraete F. , Wolf M. M.  \and Cirac J. I.}
    \REVIEW{Nat. Phys}{5}{2009}{633}.

\bibitem{Carvalho11}
    \Name{Stevenson R. N., Hope J. J. \and  Carvalho A. R. R.}
    \REVIEW{Phys. Rev. A}{84}{2011}{022332}.

\bibitem{Gisin01}
    \Name{Kwiat P. G. \etal}
    \REVIEW{Nature}{409}{2001}{1014}.
Kwiat P. G.,  Barraza-Lopez S., Stefanov A. \and Gisin N.

\bibitem{Lidar98}
    \Name{Lidar D. A.,  Chuang I. L. \and Whaley K. B.}
    \REVIEW{Phys. Rev. Lett.}{81}{1998}{2594}.

\bibitem{Kim12}
    \Name{Kim Y.-S. \etal}
    \REVIEW{Nat. Phys.}{8}{2012}{117}.


\bibitem{ManiscalcoPRL08}
    \Name{Maniscalco S. \etal}
    \REVIEW{Phys. Rev. Lett.}{100}{2008}{090503}.

\bibitem{Viola99}
    \Name{Viola L.,  Knill E.  \and  Lloyd S.}
    \REVIEW{Phys. Rev. Lett.}{82}{1999}{2417}.

\bibitem{CarvalhoNJP11}
    \Name{Carvalho A. R. R. \and Santos M. F.}
    \REVIEW{New J. Phys.}{13}{2011}{013010}.

\bibitem{Plenio02}
    \Name{Plenio M. B. \and Huelga S. F.}
    \REVIEW{Phys. Rev. Lett.}{88}{2002}{197901}.

\bibitem{Hartmann06}
    \Name{Hartmann L.,  D\"ur W. \and Briegel  H.-J.}
    \REVIEW{Phys. Rev. A}{74}{2006}{052304}.

\bibitem{Krauter11}
    \Name{Krauter  H. \etal}
    \REVIEW{Phys. Rev. Lett.}{107}{2011}{080503}.

\bibitem{Braun02}
    \Name{Braun  D.}
    \REVIEW{Phys. Rev. Lett.}{89}{2002}{277901}.

\bibitem{Benatti03}
    \Name{Benatti F.,  Floreanini R. \and  Piani M.}
    \REVIEW{Phys. Rev. Lett.}{91}{2003}{070402}.

\bibitem{Tanas08}
    \Name{Ficek Z. \and  Tana\'{s} R.}
    \REVIEW{Phys. Rev. A}{77}{2008}{054301}.

\bibitem{Tudela11}
    \Name{Gonzalez-Tudela A.  \etal}
    \REVIEW{Phys. Rev. Lett.}{106}{2011}{020501}.

\bibitem{Xu11}
    \Name{J. Xu  \etal}
    \REVIEW{Phys. Rev. A}{84}{2011}{032334}.

\bibitem{Quiroga07}
    \Name{Quiroga L. \etal}
    \REVIEW{Phys. Rev. A}{75}{2007}{032308}.

\bibitem{Huang09}
    \Name{Huang X. L., Guo J. L.  \and  Yi X. X.}
    \REVIEW{Phys. Rev. A}{80}{2009}{054301}.

\bibitem{Camalet2011}
    \Name{Camalet S.}
    \REVIEW{Eur. Phys. J. B.}{84}{2011}{467}.

\bibitem{Wu11}
    \Name{Wu L.-A. \and Segal D.}
    \REVIEW{Phys. Rev. A}{84}{2011}{012319}.

\bibitem{Znidaric12}
    \Name{\u{Z}nidari\u{c} M.}
    \REVIEW{Phys. Rev. A}{85}{2012}{012324}.

\bibitem{Camalet2013}
    \Name{Camalet S.}
    \REVIEW{Eur. Phys. J. B.}{86}{176}{2013}.

\bibitem{Linden11}
    \Name{Brunner N. \etal}
    \REVIEW{Phys. Rev. E}{85}{2012}{051117}.


\bibitem{Joulain}
    \Name{Joulain K. \etal}
    \REVIEW{Surf. Sci. Rep.}{57}{2005}{59}.

\bibitem{Messina12}
    \Name{Messina R., Antezza M. \and Ben-Abdallah P.}
    \REVIEW{Phys. Rev. Lett.}{109}{2012}{244302}.

\bibitem{AntezzaPRL05}
    \Name{Antezza M., Pitaevskii L. P. \and Stringari S.}
    \REVIEW{Phys. Rev. Lett.}{95}{2005}{113202}.

 \bibitem{AntezzaJPA06}
    \Name{Antezza M.}
    \REVIEW{J. Phys. A: Math. Gen.}{39}{2006}{6117}.

\bibitem{ObrechtPRL07}
    \Name{Obrecht J. M. \etal}
    \REVIEW{Phys. Rev. Lett.}{98}{2007}{063201}.

\bibitem{AntezzaPRA08}
    \Name{Antezza M. \etal}
    \REVIEW{Phys. Rev. A}{77}{2008}{022901}.
Antezza M., Pitaevskii L. P., Stringari S. \and  Svetovoy V. B.

\bibitem{Bimonte09}
    \Name{Bimonte G.}
    \REVIEW{Phys. Rev. A}{80}{2009}{042102}.

\bibitem{KardarPRL2011}
    \Name{Kr\"{u}ger M., Emig T. \and Kardar M.}
    \REVIEW{Phys. Rev. Lett.}{106}{2011}{210404}.

\bibitem{MesAntEPL11}
    \Name{Messina R. \and Antezza M.}
    \REVIEW{Europhys. Lett.}{95}{2011}{61002}.
\bibitem{MesAntPRA11}
    \Name{Messina R. \and Antezza M.}
    \REVIEW{Phys. Rev. A}{84}{2011}{042102}.

\bibitem{BellomoEPL2012}
    \Name{Bellomo B.,  Messina R.  \and Antezza M.}
    \REVIEW{Europhys. Lett.}{100}{2012}{20006}.

\bibitem{BellomoPRA2013}
    \Name{Bellomo B.,  Messina R., Felbacq D.  \and Antezza M.}
    \REVIEW{Phys. Rev. A}{87}{2013}{012101}.


\bibitem{CohenTannoudji97}
    \Name{Cohen-Tannoudji C., Dupont-Roc J. \and  Grynberg G.}
    \Book{Photons and Atoms: Introduction to Quantum Electrodynamics}
    \Publ{Wiley}
    \Year{1997}

\bibitem{Agarwal1974}
    \Name{Agarwal G. S.}
    \Book{\emph{in} Quantum Statistical Theories of Spontaneous Emission and their Relation to other Approaches, \emph{edited by G. H\"{o}hler, Springer Tracts in Modern Physics Vol. 70}}
    \Publ{Springer-Verlag, Berlin}
    \Year{1974}

\bibitem{FicekBook2005}
    \Name{Ficek Z. \and Swain S.}
    \Book{Quantum Interference and Coherence: Theory and Experiments}
    \Publ{Springer, New York}
    \Year{2005}

\bibitem{BellomoASL}
    \Name{Bellomo B., Lo Franco R. \and Compagno G.}
    \REVIEW{Adv. Sci. Lett.}{2}{2009}{459}.

\bibitem{Pratt2004PRL}
    \Name{Di Carlo  L.}
    \REVIEW{Nature}{460}{2009}{240}.

\bibitem{wootters1998PRL}
    \Name{Wootters W. K.}
    \REVIEW{Phys. Rev. Lett.}{80}{1998}{2245}.

\bibitem{Yu2007}
    \Name{Yu  T.  \and Eberly J. H.}
    \REVIEW{Quantum Inf. Comput.}{7}{2007}{459}.

\bibitem{Bellomopreparation}
    Bellomo B. \and Antezza M., \textit{in preparation.}

\bibitem{horodecki}
    \Name{Horodecki M., Horodecki P. \and Horodecki R.}
    \REVIEW{Phys. Rev. Lett.}{78}{1997}{574}.


\end{thebibliography}
\end{document}